\documentclass{ws-p8-50x6-00}

\begin{document}

\title{Vacuum Energy, Cosmological Supersymmetry Breaking
and Inflation 
from Colliding Brane Worlds.}

\author{N.E. Mavromatos}

\address{Department of Physics, Theoretical Physics, 
King's College London, Strand, London WC2R 2LS, United Kingdom}


\maketitle

\abstracts{In the context of 
colliding brane worlds I discuss a toy cosmological 
model, developed in collaboration with {\bf E. Gravanis}, 
which arguably produces inflation 
and a relaxing to zero cosmological ``constant''
hierarchically small as compared to the supersymmetry breaking (TeV) 
scale. Supersymmetry breaking is induced by compactification 
of the brane worlds on magnetized tori. 
The crucial ingredient is the non-criticality (non conformality) 
of string theory
on the observable brane world induced at the collision, which is thus
viewed as a cause for departure from equilibrium in this system.
The hierarchical smallness of the present-era vacuum energy, 
as compared to the SUSY breaking scale, is thus attributed to relaxation.}


One of the most  important unsolved puzzles 
in Theoretical Particle Physics
is the issue of the smallness of the Cosmological Constant
(or, better vacuum energy density) in comparison with other
physical scales, for instance, 
the scale at which Supersymmetry is broken
in supersymmetric theories. 
The resolution of such puzzles may lie in the 
way by which supersymmetry is broken. 
In a recent work~\cite{gravmav}, whose contents are reviewed
briefly here, 
we presented a (toy model) scenario, according to which
colliding branes in superstring theory may result in
a way of breaking supersymmetry on our four-dimensional
world at a TeV scale, while maintaining a very small
vacuum energy, decreasing with cosmological time (relaxation). 
The model 
consists of two five-branes of type IIB string theory, 
embedded in a ten dimensional bulk space time. 
Two of the longitudinal brane dimensions are assumed compactified
on a small torus, of 
radius $R$. In one of the branes,
from now on called {\it hidden}, the torus is {\it magnetized} 
with a 
constant magnetic
field of intensity H. This amounts to an effective 
four-dimensional vacuum energy in that brane of order:
$V_{\rm hidd} = R^2H^2 > 0$. Notice that such compactifications
provide alternative ways of breaking supersymmetry~\cite{bachas},
which we shall make use of in the current article. 
In scenaria with two branes embedded 
in higher-dimensional bulk space times it 
is natural to assume (from the point of view
of solutions to bulk field equations) that the two branes
have {\it opposite} tensions. We, therefore, assume that 
before the collision the visible 
brane (our world) has negative tension $V_{\rm vis}=-V_{\rm hidd} < 0$.
A negative tension brane is consistent with the possibility of 
accepting supersymmetric theories on it (anti-de-Sitter type).
The presence of opposite tension branes implies that 
the system is not stable, but this is O.K. from a cosmological
view point. 

For our purposes we assume that the two 
branes are originally on {\it collision course}
in the bulk, with a relative velocity $u \ll 1$ for the 
validity of the $\sigma$-model perturbation theory, and to allow
for the model to have predictive power. 
The collision takes place at a given time moment. 
This constitutes an event, which 
in our scenario is identified with the {\it initial
cosmological singularity} (big bang) on the
observable world. We note that 
similar scenaria
exist in the so-called ekpyrotic model for the Universe~\cite{ekpyrotic}.
It must be stressed, though, that 
the similarity pertains only to the brane-collision event. 
In our approach the physics is entirely different from 
the ekpyrotic scenario. The collision  
is viewed as an event resulting in non-criticality 
(departure from conformal invariance) of the underlying string theory,
and hence in non-vanishing $\beta$ functions at a $\sigma$-model level. 
On the contrary, in the scenario of \cite{ekpyrotic} the 
underlying four-dimensional effective theory (obtained after integration
of the bulk extra dimensions~\cite{ekpyrotic,linde}) is assumed always
critical, satisfying classical equations of motion, and hence vanishing 
$\sigma$-model $\beta$
functions. This latter property leads only to contracting 
and {\it not expanding} four-dimensional
Universes according to the work of \cite{linde}, which constitutes
one of the main criticisms of the ekpyrotic universe. 
On the other hand, in our non-critical description of the collision
we do not assume classical solutions of the equations of motion,
neither specific potentials associated with bulk branes, as 
in \cite{ekpyrotic}. 

The physics of our colliding worlds model can be summarized as follows: 
During the collision one assumes 
electric current transfer 
from the hidden to the visible brane, which results in 
the appearance of a magnetic field on the 
visible brane. We also assume that the entire effect
is happening very
slowly and amounts to a slow flow of energy and current
density from the positive energy density brane to
the one with negative tension. 
In turn, this results in a
positive energy component
of order $H^2R^2$  
in the vacuum energy of the visible brane world.
This energy component, together with contributions 
from recoil, may be 
assumed to {\it cancel } the pre-existing negative tension asymptotically
in time, leading to a vanishing cosmological constant at $t=\infty$. 
At the moment of the collision the conformal invariance 
of the $\sigma$-model describing (stringy) excitations on the observable
brane world is spoiled, thereby implying the need for Liouville 
dressing~\cite{ddk,emn}. This procedure restores 
conformal invariance at the cost of introducing 
an extra target space coordinate (the Liouville mode $\phi$),
which in our model has time-like signature.
Hence, initially, one
faces a two-times situation. 
We argued~\cite{gravmav}, though, that
our observable (cosmological) time $X^0$ parametrizes a certain curve, 
$\phi={\rm const.}\;X_0+{\rm const.'}$, on the
two-times plane $(X_0,\phi)$, and hence one is left with one
physical time.  

The appearance of the magnetic field on the visible 
brane, on the dimensions $X^{4,5}$, 
is described (for times long after the collision)
within a $\sigma$-model
superstring formalism by the boundary deformation:
${\cal V}_{H}= \int _{\partial \Sigma} 
A_5\partial_\tau X^5 + {\rm supersymm.~partners}$,
where $A_5=e^{\varepsilon X^0} H X^4$, and 
$\partial_\tau$ denotes tangential $\sigma$-model derivative
on the world-sheet boundary.
This $\sigma$-model 
deformation 
describes open-string excitations attached to the brane world
(c.f. fig. 1).
The presence of the quantity 
$\varepsilon \to 0^+$
reflects the {\it adiabatic} switching 
on of the magnetic field
after the collision. It should be remarked that in our approach 
the quantity $\varepsilon$ is viewed as a world-sheet 
renormalization-group scale parameter~\cite{gravmav}.  
In addition to the magnetic field deformation, the $\sigma$-model 
contains also boundary deformations describing the `recoil' of the 
visible world due to the collision:
${\cal V}_{\rm rec} = \int _{\partial \Sigma}  Y_6(X_0)\partial _n X^6
+{\rm supersymm.~partners}$,
where $Y_6(X_0)=u X^0 e^{\varepsilon X^0}$,
$\partial_n$ denotes normal $\sigma$-model derivative
on the world-sheet boundary, and $u$ is 
the recoil velocity of the visible brane world.

\begin{figure}[htb]
\begin{center}
\epsfxsize=3in
\bigskip
\centerline{\epsffile{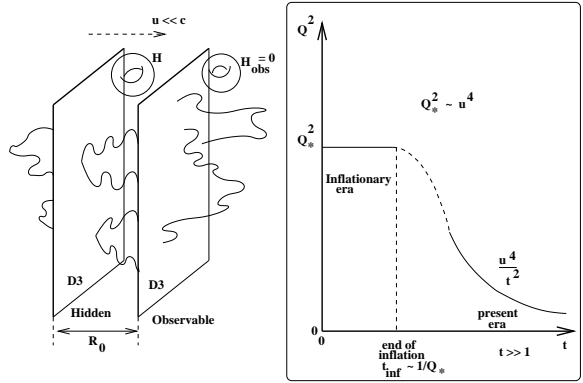}}
\caption{{\it A scenario 
involving two colliding type-II  5-branes which provides 
inflation and a 
relaxation model for cosmological vacuum energy.}}
\end{center}
\label{infla} 
\end{figure}

The presence of the exponential 
factors $e^{\varepsilon X^0}$ in 
{\it both} the magnetic field and recoil 
deformations 
implies a small but negative 
world-sheet anomalous dimension
$-\frac{\varepsilon ^2}{2} < 0$, and hence the relevance of 
both operators from a renormalization-group
point of view. The induced central-charge deficit $Q^2$,
which quantifies the departure from the conformal point of the 
pertinent $\sigma$-model, can be 
computed
by virtue of the Zamolodchikov's c-theorem~\cite{gravmav}:
$\frac{d}{d {\cal T}}Q^2 \sim - \frac{H^2 + u^2}{{\cal T}^2} \quad \to
\quad 
Q^2 ({\cal T}) = Q_0^2 + \frac{H^2 + u^2}{{\cal T}}~,~{\cal T}
={\rm ln}|L/a|^2~$, with $L (a)$ the 
world-sheet infrared (ultraviolet) cutoff scale.
As discussed in detail in 
\cite{gravmav}, the correct scaling behaviour
of the operators necessitates the 
identification ${\cal T} \sim \varepsilon^{-2}$ which we assume from now on. 
The quantity 
$Q_0^2 = Q^2 (\infty)$  
is the equilibrium
vacuum energy density, which we take 
to be zero $Q_0^2 =0$ due to the cancellation between the initial
vacuum energies of the colliding branes assumed in our model. 

The non-conformal deformed $\sigma$-model can become conformal
as usual by Liouville dressing~\cite{ddk,gravmav}:
$
A_5(X_0,X_4,\phi)=H X^4e^{\varepsilon X^0+ \alpha \phi} , \;\; 
Y_6(X_0,\phi)=u X^0 e^{\varepsilon X^0+ \alpha \phi}$,
where $\alpha \sim \varepsilon$ are the Liouville anomalous 
dimensions~\cite{ddk} and $\phi$ is the normalized Liouville mode,
whose zero-mode is related to the renormalization-group scale ${\cal T}$ 
as: $\phi = Q({\cal T}){\cal T}$. From 
the work of \cite{bachas} it becomes clear that the coupling constant 
$H$ is associated with supersymmetry-breaking mass splittings.
This has to do with the different way fermions and bosons couple to an 
external magnetic field. The mass splittings squared of an open string 
are in our case~\cite{gravmav}: 
\begin{equation} 
\Delta m_{\rm string}^2 \sim   He^{\alpha \phi + 
\varepsilon X^0 }\Sigma_{45} 
\label{split}
\end{equation}
The so-obtained mass splittings are constant upon the requirement 
that the flow of time $X^0$ and of Liouville mode $\phi$ are correlated
in such a way that 
\begin{equation} 
\varepsilon X^0 + \varepsilon \phi/\sqrt{2} 
={\rm constant}~,
\label{liouvtime}
\end{equation}
or at most slowly varying.
Notice that
deviations from the condition (\ref{liouvtime}) 
would result in very
large negative-mass squares, which are clearly unstable configurations.
Hence, the identification (\ref{liouvtime}) seems to provide
a resolution of this problem.
To ensure the phenomenologically reasonable order of magnitude of a TeV scale,
one must assume very small~\cite{bachas} $H \sim 10^{-30} \ll 1$ 
in Planck units.
Note also that parametrizing the condition (\ref{liouvtime})  
as $X^0=t$, $\phi_0 =\sqrt{2}t$,
and taking into account that, for convergence of $\sigma$-model 
path integration, it is formally necessary to work with Euclidean signature 
$X^0$~\cite{gravmav}, 
the induced 
metric on the hypersurface (\ref{liouvtime}) in the 
extended space time  
acquires a 
Minkowskian-signature Robertson-Walker form: $ds^2_{\rm hypersurf} = -(d\phi_0)^2 + (dX^0)^2 + \dots = - (dt)^2 + \dots~$. 
where $\dots$ denote spatial parts. In \cite{gravmav}, where we refer the 
interested reader,
we have given some arguments on a 
a {\it dynamical stability} of the condition (\ref{liouvtime}) in the 
context of Liouville strings. 
Physically, one may interpret this 
result as 
implying that a time-varying magnetic field 
induced by the collision implies
back reaction of strings onto the space time 
in such a way that the mass splittings of the string excitation
spectrum, as a result of the field, are actually stabilized.

We now notice that in our case the dilaton field is 
$\Phi = Q\phi = Q^2 \varphi \sim (H^2 + u^2)$,
that is, one faces a situation with an asymptotically 
constant dilaton. 
This is a welcome fact, because otherwise, the space-time
would not be asymptotically flat, and one could face trouble in
appropriately defining masses. In the case of a constant dilaton the 
vacuum energy is determined by the central-charge charge 
deficit $Q^2$, which in our case is: 
\begin{equation}
\Lambda = \frac{R^{2n}}{\phi_0^2}(H^2 + u^2)^2
\label{cosmoconst}
\end{equation}
where $\phi_0=t$ is the world-sheet zero mode of the
Liouville field to be identified with the target time
on the {\it hypersurface} (\ref{liouvtime}) of the extended
space-time resulting after Liouville dressing. 

We next remark that the restoration of the conformal 
invariance by the Liouville mode results in the following equations
for the $\sigma$-model background fields/couplings 
$g^i$ near a fixed-point of the world-sheet renormalization group
(large-times cosmology) we restrict ourselves 
here~\cite{ddk}:
$(g^i)'' + Q (g^i)' = -\beta^i (g)$,
where the prime denotes derivative with respect to the 
Liouville zero mode $\phi_0$, and the sign 
on the right-hand-side is appropriate for supercritical 
strings we are dealing with here. 
These equations replace the equilibrium equations $\beta^i=0$ 
of critical string theory, and should be used in our colliding brane scenario
to determine the evolution of the scale factor of the four-dimensional 
Robertson-Walker Universe. A preliminary analysis has been performed
in \cite{gravmav}, where we refer the reader for details. 

Below we only describe briefly the main results. 
In our non-critical string scenario,
one does indeed obtain an expanding Universe, in contrast to 
standard ekpyrotic scenaria~\cite{ekpyrotic,linde}, based on 
critical strings and specific solutions to classical equations 
of motion. 
One of the most important features of the existence of a non-equilibrium 
phase of string theory due to the collision is the possibility
for an {\it inflationary phase}. Although the physics near the collision
is strongly coupled, and the $\sigma$-model perturbation theory
is not reliable, nevertheless one can give compelling 
physical arguments favoring the existence of an
early phase of the brane world where the four-dimensional 
Universe 
scale factor undergoes  exponential growth (inflation). 
This can be understood as follows: 
in our model we encounter two type-II string theory branes colliding, and then 
bouncing back. From a stringy point of view the collision and bounce 
will be described by a phase where open strings stretch between the 
two branes worlds (which can be thought of as lying a few string scales
apart during the collision, c.f. fig. 1). During that early 
phase the excitation
energy of the brane worlds can be easily computed by the same methods
as those used to study scattering of type II D-branes in \cite{bachas2}.
In type II strings the exchange of pairs of open strings
is described by annulus world-sheet diagrams, which in turn 
results in the appearance of ``spin structure factors'' 
in the scattering amplitude. The latter are expressed in terms of 
appropriate sums over Jacobi $\Theta$ functions. 
Due to special properties of these functions, 
the 
spin structures start of at quartic order in $u$~\cite{bachas2}.
The resulting excitation energy is therefore 
of order ${\cal O}(u^4)$ and may be thought 
off as an initial value of the central charge deficit of the 
non-critical string theory describing the 
physics of our brane world after the collision. 
The deficit $Q^2$ 
is 
thus cut off at a finite value in the (world-sheet) 
infrared scale (early target times, c.f. fig. 1). 
One may plausibly assume that the central charge deficit remains constant
for some time, which is the era of {\it inflation}. 
It can be shown~\cite{gravmav} that for (finite) 
constant 
$Q^2=Q_*^2= {\cal O}(u^4)$ the Liouville equations imply 
a scale factor exponentially growing with the Liouville zero mode  
$a(\phi_0) = e^{Q_*\phi_0/2}$. Upon the condition (\ref{liouvtime}), then, 
one obtains an early inflationary phase after the collision, in contrast 
to 
the critical-string based arguments of \cite{linde}. The duration of 
the inflationary phase is $t_{\rm inf} \sim 1/Q_* \sim {\cal O}(u^{-2})$,
which yields the conventional values of inflationary models
of order $t_{\rm infl} \sim 10^9 t_{\rm Planck}$ for $u^2 \sim 
10^{-9}$. This concludes our discussion on this toy model.

\section*{Acknowledgements} 

I wish to thank A. Faraggi and S. Abel for the invitation. 
This work is partially supported by the European Union
(contract ref. HPRN-CT-2000-00152).

\end{document}